\documentclass[12pt]{spieman}  
\usepackage{amsmath,amsfonts,amssymb}
\usepackage{graphicx}
\usepackage{setspace}
\usepackage{tocloft}
\usepackage{lineno}
\usepackage{siunitx}
\title{Off-axis performance characterization of hard X-ray optics for the FOXSI-4 and FOXSI-5 sounding rocket missions}

\author[a]{Yusuke Yoshida}
\author[a]{Kazuki Ampuku}
\author[b]{Koki Sakuta}
\author[a]{Makoto Yoshihara}
\author[a]{Keitoku Yoshihira}
\author[a]{Ryoma Tanaka}
\author[a]{Tetsuo Kano}
\author[a]{Naoki Ishida}
\author[a]{Wataru Kato}
\author[c]{Yoshitaka Inoue}
\author[d,e]{Keisuke Tamura}
\author[d]{Takashi Okajima}
\author[f]{Kikuko Miyata}
\author[g]{Noriyuki Narukage}
\author[h]{Gota Yamaguchi}
\author[h]{Hidekazu Takano} 
\author[h]{Yoshiki Kohmura} 
\author[i]{Shutaro Mohri}
\author[i]{Satoru Egawa}
\author[i]{Hiroto Motoyama}
\author[j]{Takehiro Kume}
\author[j]{Yusuke Matsuzawa}
\author[j]{Yoichi Imamura}
\author[j]{Takahiro Saito}
\author[j]{Kentaro Hiraguri}
\author[j]{Hirokazu Hashizume}
\author[i, h]{Hidekazu Mimura}
\author[a]{Ikuyuki Mitsuishi}
\affil[a]{Nagoya University, Furo-cho, Chikusa-ku, Nagoya, Japan}
\affil[b]{Tohoku University, 6-3 Aoba, Aramaki, Aoba-ku, Sendai, Japan}
\affil[c]{IMV CORPORATION, 3-24-5 Jojo-cho, Kasugai, Japan}
\affil[d]{NASA/GSFC, 8800 Greenbelt Rd, Greenbelt, United States}
\affil[e]{University of Maryland, 7901 Regents Drive, College Park, United States}
\affil[f]{Meijo University, 1-501 Shiogamaguchi, Tempaku-ku, Nagoya, Japan}
\affil[g]{National Astronomical Observatory of Japan, 2-21-1 Osawa, Mitaka, Japan}
\affil[h]{RIKEN/SPring-8, 1-1-1, Kouto, Sayo-cho, Sayo-gun, Hyogo, Japan}
\affil[i]{The University of Tokyo, 4-6-1 Komaba, Meguro-ku, Tokyo, Japan}
\affil[j]{Natsume Optical Corporation, 3461 Kamichaya, Kanae, Iida, Japan}

\cftpagenumbersoff{figure}
\cftpagenumbersoff{table} 
\begin{document} 
\maketitle

\begin{abstract}
FOXSI-4 and FOXSI-5 are Japan--U.S. joint sounding rocket missions investigating particle acceleration and plasma heating in solar flares.
We developed soft and hard X-ray optics for both missions and characterized the off-axis response of the hard X-ray optics.
Imaging performance was measured at SPring-8 BL29XUL, and effective areas at NASA/GSFC.
With increasing off-axis angle, focused images became elongated and the large-angle component of the point spread function increased.
The Half Power Diameter (HPD) remained below 20 arcsec up to 7.5 arcmin, increasing to approximately 50 arcsec at 15 arcmin.
A Wolter-I optics model reproduced the off-axis HPD dependence within approximately 20\%, with residuals likely associated with azimuthal variations in mirror figure errors.
On-axis effective areas were 41.3 $\pm$ 1.9 mm$^2$ at 6.4 keV and 32.3 $\pm$ 1.6 mm$^2$ at 10 keV, decreasing by 20\% and 30\%, respectively, at 5 arcmin.
Incorporating an empirical correction for geometric effects and X-ray reflectivity reproduced measured effective areas within 20\% over 1.0--12.5 keV up to 5 arcmin.
At larger off-axis angles, deviations at lower energies suggest increasing stray-light contributions.
These measurements provide off-axis response data for constructing response functions for FOXSI-4 and FOXSI-5 analyses.
\end{abstract}

\keywords{X-rays, X-ray optics, Calibration, Sounding rockets, Sun, Solar flare}

{\noindent \footnotesize\textbf{*}Ikuyuki Mitsuishi,  \linkable{mitsuisi@u.phys.nagoya-u.ac.jp} }

\begin{spacing}{2}   

\section{Introduction}
\label{sect:intro}  
The Sun is the closest star to the Earth and provides detailed information on astronomical phenomena in time, space, and energy. Many of the physical phenomena observed on the Sun are fundamental and universal, and the knowledge gained from solar observations can be applied not only to solar physics but also to a wide range of fields, including astrophysics and plasma physics\cite{Takasao/AAS/2020, Ampuku/AAS/2026, Zweibel/Annual_Review/2009, Guo/SSR/2024}. In particular, in the solar corona and flares, energy release associated with magnetic reconnection leads to particle acceleration and plasma heating. Therefore, observationally clarifying these processes is important for understanding the mechanisms of energy conversion in cosmic plasmas. 
Magnetic reconnection in the solar corona accelerates electrons in and around the reconnection region above flare loop tops. These accelerated electrons produce nonthermal hard X-ray emission through bremsstrahlung near the loop tops and at the footpoints, where they deposit their energy into the chromosphere. This energy deposition drives chromospheric evaporation, filling the flare loops with hot thermal plasma that emits soft X-rays through thermal bremsstrahlung and line emission\cite{Masuda/PASJ/1995,Khan/AA/2004}. 
Therefore, X-rays provide an important means of investigating particle acceleration and plasma heating in the solar corona and flares, and these physical processes have been extensively studied using various X-ray observatories\cite{Bohlin/Solar.Physics/1980,Tanaka/Solar.Physics/1983,Ogawara/Solar.Physics/1991,Lin/Solar.physics/2002,Kosugi/Solar.Physics/2007,Muller/AA/2020}. The X-ray Telescope (XRT) \cite{Golub/Solar.Physics/2007} aboard \textit{Hinode}, launched in 2006 \cite{Kosugi/Solar.Physics/2007}, achieved direct imaging with high spatial resolution in the soft X-ray band. However, XRT does not support imaging spectroscopy that directly measures the energy of individual X-ray photons. The \textit{Reuven Ramaty High Energy Solar Spectroscopic Imager} (\textit{RHESSI}) \cite{Lin/Solar.physics/2002}, launched in 2002, provided high spatial resolution and spectroscopic capabilities in the hard X-ray band. However, because it adopted an indirect imaging technique, its sensitivity and dynamic range were limited. Thus, imaging spectroscopy in the soft X-ray band and a high dynamic range capable of simultaneously resolving bright hard X-ray sources at footpoints from faint hard X-ray sources in the corona remained important challenges. If these challenges are addressed, it will be possible to capture the spatial and energy distributions of hot thermal plasmas and nonthermal electrons in solar flares in detail, providing an opportunity to observationally clarify how magnetic energy is converted into particle acceleration and plasma heating. 

Under these circumstances, the Focusing Optics X-ray Solar Imager (FOXSI), a Japan--U.S. joint sounding-rocket experiment aimed at understanding high-energy physical phenomena in the solar corona, has demonstrated imaging spectroscopy of solar soft and hard X-rays using direct focusing optics\cite{Krucker/SPIE.proc/2013, Steven/JAI/2016, Musset/SPIE.proc/2019}. FOXSI carried seven X-ray telescopes employing direct focusing optics, and the first three flights observed solar active regions, microflares, and the quiet Sun.\cite{Krucker/AAS/2014, Ishikawa/Nature.Astronomy/2017,Athiray/AA/2020,Vievering/AA/2021, Milo/AA/2022}. The fourth flight, FOXSI-4, aimed to investigate particle acceleration and plasma heating mechanisms associated with solar flares. A trigger observation strategy was planned, in which the sounding rocket was launched after the occurrence of a flare satisfying threshold conditions\cite{Buitrago-Casas/SPIE.proc/2021}. To spatially resolve flare structures in detail, FOXSI-4 was equipped with seven X-ray telescopes, including newly fabricated high-angular-resolution X-ray optics\cite{Baumgartner/SPIE.proc/2023,Ampuku/SPIE.proc/2024}. These X-ray optics were coupled with two CMOS detectors for soft X-ray observations\cite{Shimuzu/SPIE.proc/2024}, four wide-gap CdTe double-sided strip detectors (CdTe-DSD)\cite{Nagasawa/NIM/2025, Nagasawa/NIM/2023} and one Quad-Timepix3 CdTe detector\cite{Anna/SPIE.proc/2026} for hard X-ray observations. FOXSI-4 was launched from Poker Flat Research Range, Alaska, on April 17, 2024, at 22:13:00 (UTC), and successfully achieved the first soft and hard X-ray imaging spectroscopy observation of an M-class flare using direct focusing optics, coordinated with the Hi-C Flare observation\cite{Tiwari/AAS/2025}. The re-flight mission FOXSI-5 was launched on May 14, 2026, at 19:23:30 (UTC), and successfully observed three active regions following C-class flares\cite{Buitrago/NASA.proposal/2023}.

For both FOXSI-4 and FOXSI-5, we were responsible for the development of one soft X-ray optical system and one hard X-ray optical system among the seven X-ray optics. We have conducted environmental tolerance tests such as vibration tests\cite{Fujii/SPIE.proc/2024}, ground calibration tests\cite{Ampuku/SPIE.proc/2024, Fujii/PASP/2026}, and the development of our own ray-tracing simulator to construct a response function\cite{Yoshida/SPIE.proc/2024}. Since solar flares have extended structures, it is necessary to understand not only the
response of the optics to on-axis rays but also the response to off-axis rays when analyzing observation data. For the FOXSI-4 soft X-ray optics, the off-axis response of the imaging performance and effective area has already been evaluated. Using these data, the angular misalignment between the components was estimated, and the energy dependence of the effective area for off-axis rays was reproduced with an accuracy of $\pm10\%$\cite{Ampuku/SPIE.proc/2024}. However, although the on-axis imaging performance of the hard X-ray optics onboard FOXSI-4 and FOXSI-5 has been evaluated in detail\cite{Fujii/PASP/2026}, their off-axis imaging performance and effective area have not yet been sufficiently characterized. 
In this paper, building on our earlier SPIE proceedings paper\cite{Yoshida/SPIE_proc/2026}, we present the results of ground calibration experiments for the hard X-ray optics conducted at SPring-8 and NASA Goddard Space Flight Center (GSFC), focusing on the off-axis angle dependence of the imaging performance and effective area.

\section{Our Hard X-ray Optics Onboard FOXSI-4 and FOXSI-5}
\label{sec:optics}
We have been developing space X-ray optics by applying high-precision electroforming technology originally developed for X-ray microscopes at synchrotron radiation facilities\cite{Mimura/RSI/2018}. 
The optics employ a single, monolithic, full-circumference Wolter-I mirror made of Ni, with no coating applied to its surface. 
The mirror has a focal length of 2000 mm, a diameter of 60 mm, a total length of 220 mm, and a thickness of 2 mm\cite{Yamaguchi/RSI/2023}. 
In this design, the critical energy is approximately $16$ keV for on-axis rays. Because FOXSI-4 and FOXSI-5 are designed to observe solar flares, this design prioritizes angular resolution over effective area. We integrated the mirrors into Mirror Module Assemblies (MMAs) and provided one soft X-ray optical system and one hard X-ray optical system for each mission. Details of the soft X-ray optics are presented in Ampuku et al. (2024) \cite{Ampuku/SPIE.proc/2024}, while details of the hard X-ray optics are presented in Fujii et al. (2026) \cite{Fujii/PASP/2026}. The hard X-ray optics considered in this paper consist of a front light blocker, a front aperture, a mirror cell, a mounting ring, a mirror, a rear aperture, and a rear light blocker, listed in order from the X-ray entrance side (right panel of Figure \ref{fig:Setup}). The mirror is bonded to the rear aperture. 
The hard X-ray telescope is designed to operate in the $4$--$20$ keV energy band, although its effective area decreases significantly above approximately 16 keV due to the energy dependence of the mirror reflectivity.
We evaluated the imaging performance of the hard X-ray optics for on-axis rays at SPring-8 BL29XUL. The results demonstrated an extremely sharp core with a Full Width at
Half Maximum (FWHM) of 0.7 arcsec and a Half Power Diameter (HPD) of 14 arcsec at $12$ keV under full-aperture illumination\cite{Fujii/PASP/2026}. Furthermore, a
positive correlation was found between the angular resolution and the axial direction figure error of both the primary and secondary sections, indicating that the axial figure error contributes to image degradation\cite{Fujii/PASP/2026}. 
After the FOXSI-4 flight, additional bonding was applied to the hard X-ray optics to improve their vibration tolerance. No significant change in the HPD near the optical axis was observed after the flight. Subsequent vibration tests confirmed that the optics met the required specifications, and the same optics were reflown on FOXSI-5. The results presented in this paper are therefore applicable to both missions, although differences in the alignment between the mirror optical axis and the MMA mechanical axis, as well as between the optics and the detector plane, should be considered when constructing mission-specific response functions.

\section{Ground Calibration}
\label{sec:calibration}
\begin{figure}[htbp]
\centering
\includegraphics[width=\columnwidth]{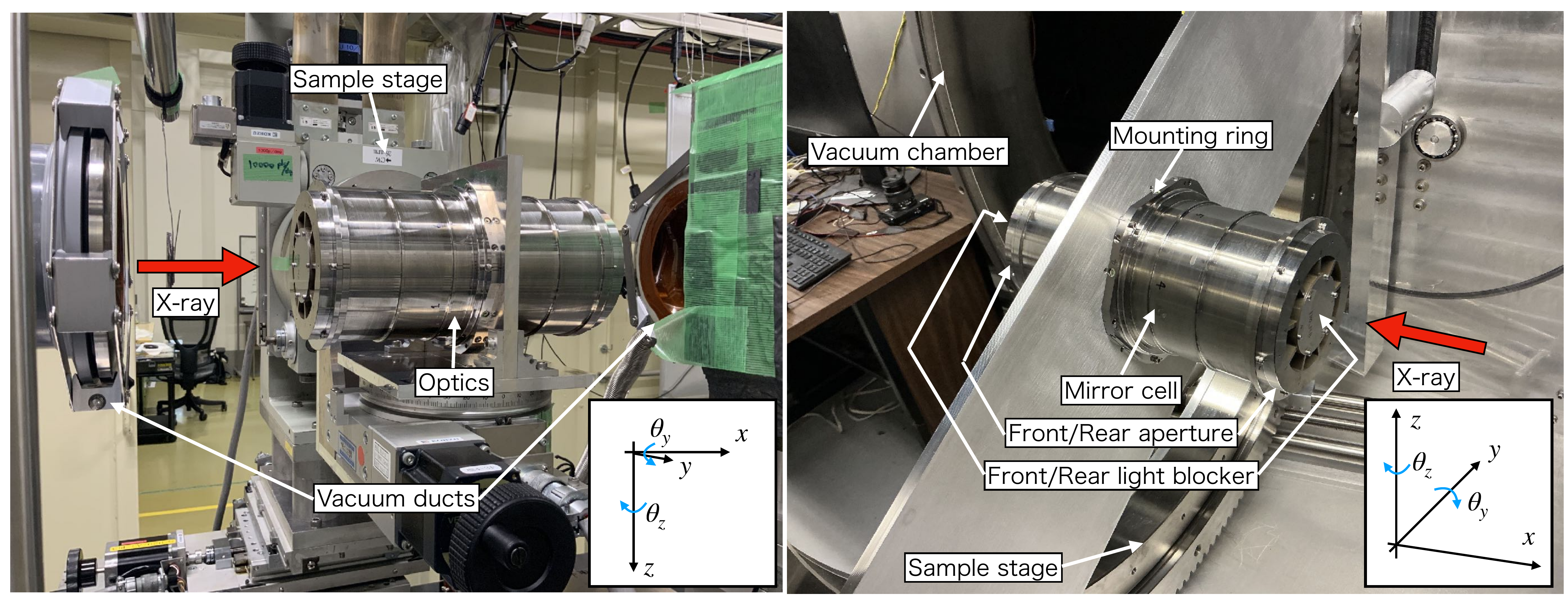}
\caption{Overview of the experimental setups at the 1 km beamline BL29XUL at SPring-8 (left) and the 100 m beamline of NASA/GSFC (right).}
\label{fig:Setup}
\end{figure}

To obtain the data required for constructing the response functions, ground calibration tests were performed. These tests were carried out at the 1 km beamline BL29XUL at SPring-8 and the 100 m beamline at NASA/GSFC, where the imaging performance and effective area for on- and off-axis rays were evaluated. An overview of the experimental setups at each facility is shown in Figure \ref{fig:Setup}. In this section, we present the ground calibration results for the hard X-ray optics.

\subsection{Imaging Performance}
\label{subsec:SP8}
The measurement was performed using the High-Brilliance X-ray Kilometer-long Large-Area Expanded-beam Evaluation System (HBX-KLAEES)\cite{Fujii/PASP/2026}, which enabled full-aperture X-ray illumination with a very small divergence angle. This capability allowed us to evaluate the imaging performance with high fidelity, resolving both the sharp core and the large-angle components of the Point Spread Function (PSF)\cite{Fujii/PASP/2026}. The coordinate system was defined as shown in the left panel of Figure \ref{fig:Setup}. The optics were mounted on translation and rotation stages with five degrees of freedom ($x$, $y$, $z$, $\theta_y$ and $\theta_z$), while the detector system was mounted on translation stages along the $y$- and $z$-axes. The detector system consisted of a scintillator, an objective lens, and a CMOS camera. The details of the experimental setup are described in Fujii et al. (2026)\cite{Fujii/PASP/2026}.

The imaging performance measurement was conducted as follows. First, the optical axis and focal position were adjusted using the same method as described in Fujii et al. (2026)\cite{Fujii/PASP/2026}. The best-aligned position was defined as the angular position that maximized the effective area and the focal position that provided the best imaging performance. Then, with the best-aligned position as the reference, $\theta_y$ and $\theta_z$ were varied from $-12.5$ arcmin to $+12.5$ arcmin in steps of $2.5$ arcmin, yielding a total of 121 angular positions at $12$ keV. Finally, images were acquired at $10$ and $15$ keV at the best-aligned position. For all of the measurements described above, dark frames were acquired under the same conditions with the upstream beam shutter closed and subtracted from the illuminated images to correct for the background. To account for temporal variations in the background, we evaluated its time dependence using a source-free region in the field of view and incorporated this variation into the background estimation.

\begin{figure}[htbp]
\centering
\includegraphics[width=0.8\columnwidth]{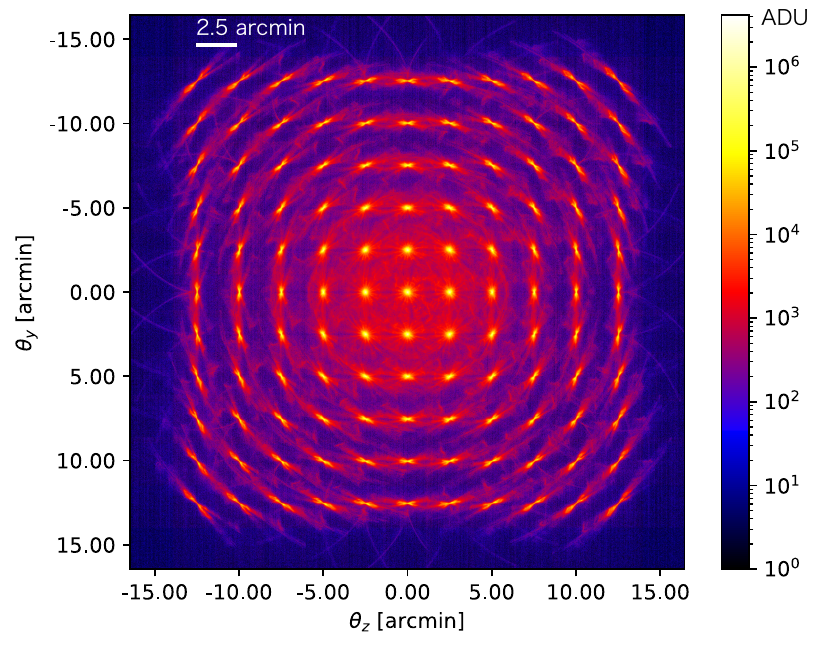}
\caption{Combined focused images as a function of off-axis angles $\theta_y$ and $\theta_z$, measured at 12 keV using HBX-KLAEES at SPring-8 BL29XUL.}
\label{fig:matrix}
\end{figure}

\begin{figure}[htbp]
\centering
\includegraphics[width=0.95\columnwidth]{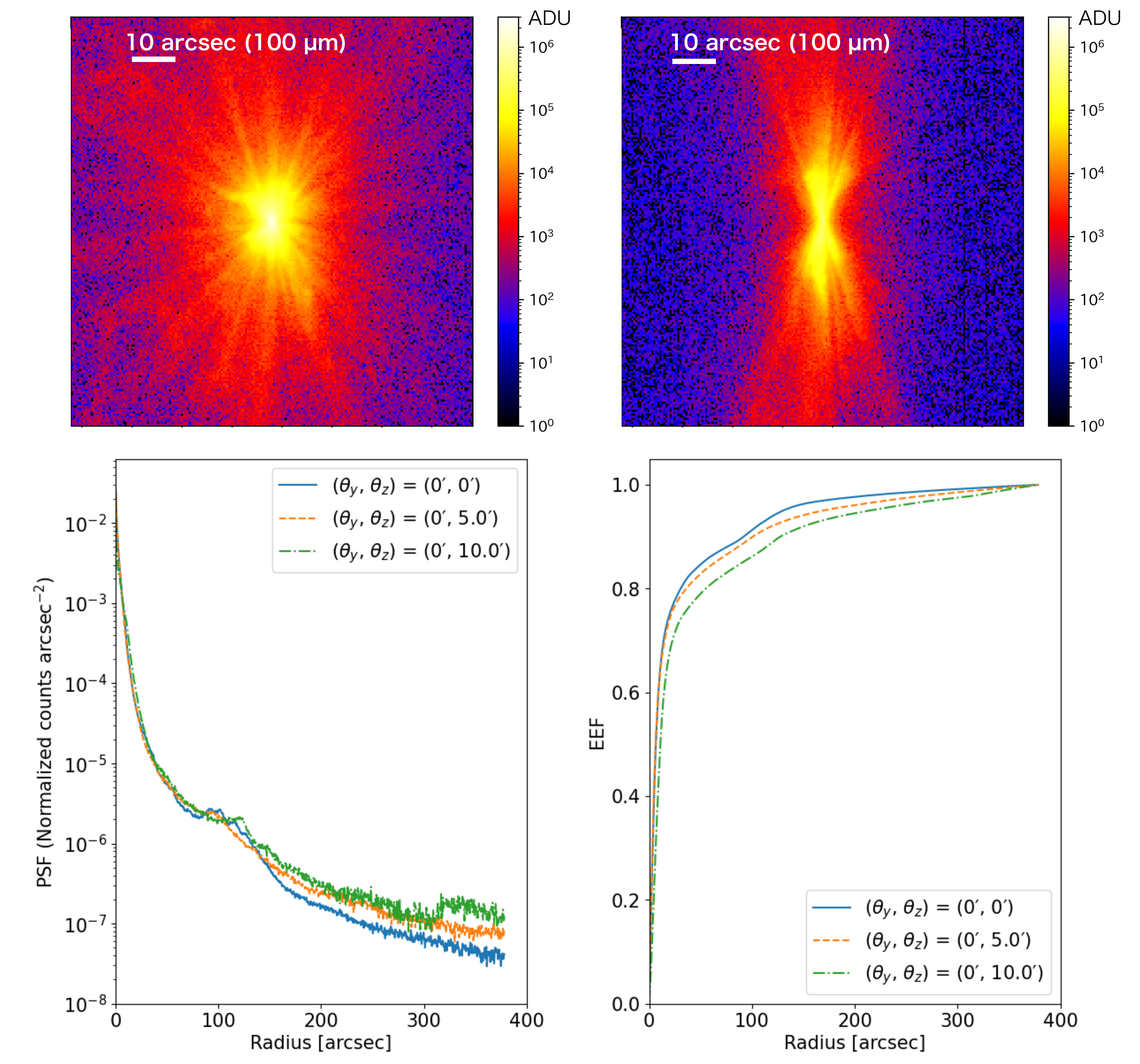}
\caption{Focused images and corresponding PSFs and EEFs at 12 keV. The upper-left and upper-right panels show the focused images at $(\theta_y,\, \theta_z) = (0',\,5.0')$ and $(0',\,10.0')$, respectively, selected from Figure \ref{fig:matrix}. The lower-left and lower-right panels show the corresponding PSFs and EEFs, respectively, calculated from these images. The on-axis image is shown in Figure 6 of Fujii et al. (2026). For comparison, the PSF and EEF for on-axis rays are also shown in the lower-left and lower-right panels, respectively. The data for $(\theta_y,\, \theta_z) = (0',\,0'),\,(0',\,5.0'),$ and $(0',\,10.0')$ are shown by blue solid, orange dashed, and green dash-dotted lines, respectively.}
\label{fig:SPring-8_psf_eef}
\end{figure}

\begin{figure}[htbp]
\centering
\includegraphics[width=0.75\columnwidth]{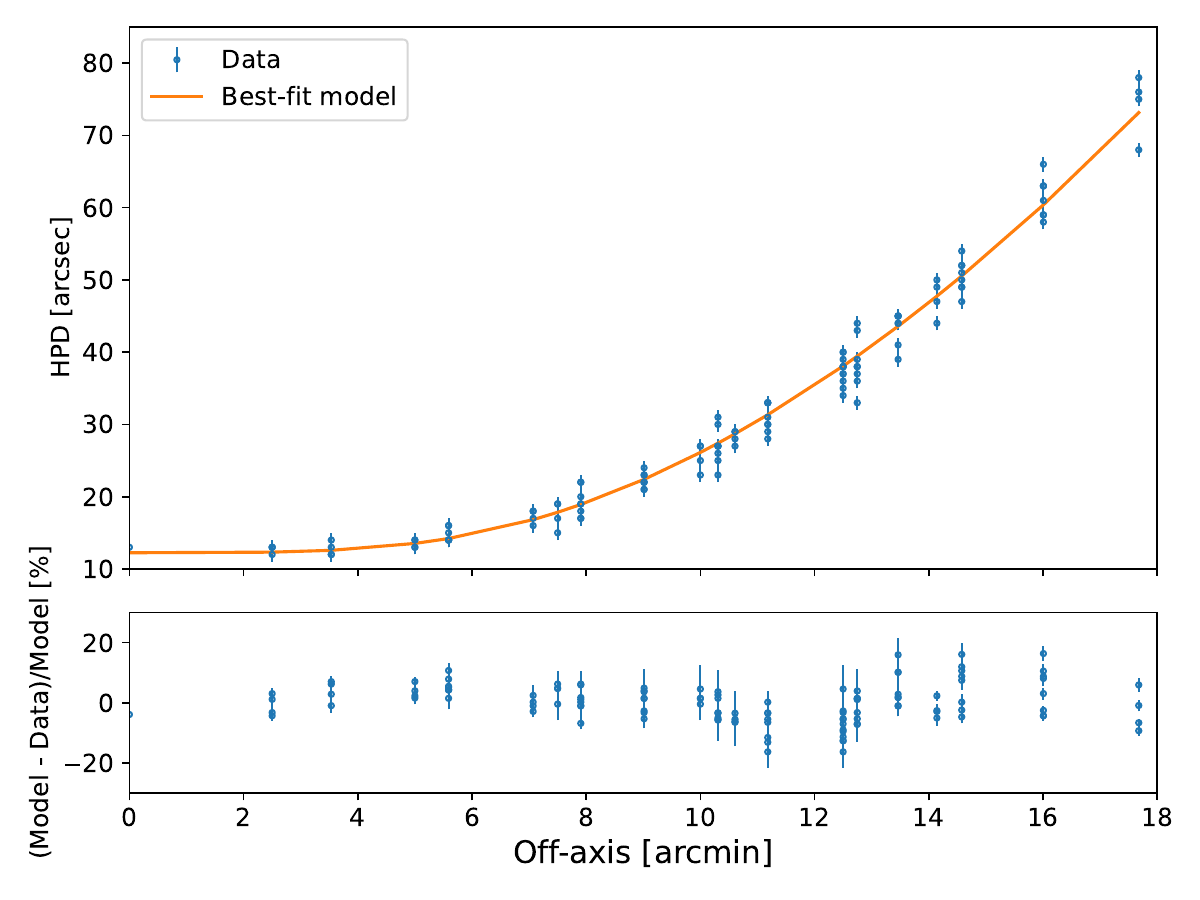}
\caption{Off-axis angle dependence of the HPD at $12$ keV for all data shown in Figure \ref{fig:matrix}. The horizontal axis represents the absolute value of the off-axis angle, defined as $\theta = \sqrt{\theta_y^2 + \theta_z ^2}$, and the vertical axis represents the HPD in arcsec. The measured HPD values are shown as blue circles, and the best-fit model using Equation \ref{eq:HPD_off-axis} is shown by the solid orange line.}
\label{fig:off-axis_hpd}
\end{figure}

The combined focused images at $12$ keV for on- and off-axis rays are shown as a function of off-axis angle in Figure \ref{fig:matrix}. 
As the off-axis angle increased, the focused images became more elongated and their brightness decreased, primarily due to the geometric characteristics of Wolter-I optics and the angular dependence of X-ray reflectivity. 
Azimuthal asymmetries were also observed, partly due to geometric obscuration by the non-axisymmetric spoke structures of the front and rear apertures, as well as mirror figure errors.
As representative examples, the focused images at $(\theta_y, \, \theta_z) = (0',\,5.0')$ and $(0',\,10.0')$ are shown in the upper panels of Figure \ref{fig:SPring-8_psf_eef}. 
The PSFs and Encircled Energy Functions (EEFs) calculated from these images are shown in the lower panels of Figure \ref{fig:SPring-8_psf_eef}. For comparison, the PSF and EEF at $(\theta_y, \, \theta_z) = (0',\,0')$ are also plotted.
Considering the size of the sensitive area of the scintillator, the integration radius for the PSF and EEF was conservatively set to 800 pixels ($\sim$ \ang{;;379}). 
Extending the integration radius to 850 pixels ($\sim$ \ang{;;402}) changed the HPD by only approximately 1 arcsec at a few positions, confirming that the choice of integration radius has little effect on the results.
The changes in the images described above are more clearly visible in the upper panels of Figure \ref{fig:SPring-8_psf_eef}. The PSF and EEF profiles also show that, as the off-axis angle increases, the intensity varies more gradually over angular scales of several tens of arcseconds, while the contribution of the large-angle component increases. 
The peak structure around 100 arcsec in the on-axis PSF, also reported in Figure 6 of Fujii et al. (2026)\cite{Fujii/PASP/2026}, is likely associated with regions of larger figure errors near the mirror edges. 
The azimuthal dependence of these figure errors is considered to cause variations in the peak intensity and shape with off-axis angle. 
In addition, the enhancement of the PSF beyond 320 arcsec at $(\theta_y,\, \theta_z) = (0',\,10.0')$ is attributed to stray light entering the integration region.
The FWHM values calculated from the PSFs were $1.9\pm0.9$ arcsec for all three positions, indicating that the sharpness of the image core remains unchanged even at an off-axis angle of 10 arcmin. However, these values correspond to the detection limit imposed by the detector pixel size. The HPD values were $13\pm1$ arcsec at $(\theta_y,\, \theta_z) = (0',\,0')$, $13\pm1$ arcsec at $(\theta_y,\, \theta_z) = (0',\,5.0')$, and $23\pm1$ arcsec at $(\theta_y,\, \theta_z) = (0',\,10.0')$. Although the degradation at an off-axis angle of 5 arcmin is small, the HPD degrades by approximately a factor of two at 10 arcmin. This degradation at larger off-axis angles is considered to be caused by increasing image blur due to the off-axis imaging characteristics of Wolter-I optics, together with the influence of stray light. The same analysis was applied to all data in the combined images, and the off-axis angle dependence of the HPD is summarized in Figure \ref{fig:off-axis_hpd}. Here, the off-axis angle is defined as $\theta = \sqrt{\theta_y^2 \, + \, \theta_z^2}$. This figure shows that the HPD remains below 20 arcsec for off-axis angles of up to 7.5 arcmin, whereas it degrades rapidly at larger off-axis angles. 
Even at the same absolute off-axis angle, the HPD exhibited azimuthal variations, with the largest difference between the maximum and minimum HPD values reaching approximately 25\%. 
An examination of the corresponding PSFs and EEFs suggests that these variations primarily arise from azimuthal differences in mirror figure errors, which affect the PSF of the double reflection component depending on the direction of incidence.

To model the off-axis angle dependence of the HPD, we assumed that the degradation in imaging performance for off-axis rays
consists of two components: the degradation caused by the off-axis imaging characteristics of Wolter-I optics and
the degradation caused by mirror figure errors suggested in Fujii et al. (2026)\cite{Fujii/PASP/2026}. L. P. VanSpeybroeck and R. C. Chase (1972)\cite{VanSpeybroeck/OPG/1972} approximated the resolution (rms blur circle radius) of a Wolter-I telescope with an optimally curved focal surface in terms of the design parameters by a simple empirical formula, as given in their Equation (2). The effect of using a flat focal plane is to double the first $\tan^2{\theta}$ term. Based on this relation, we expressed the contribution of the off-axis imaging characteristics of Wolter-I optics to the angular resolution using the following equation, where the coefficients of the first and second terms were treated as fitting parameters, $a$ and $b$, respectively.
 \begin{equation}
     \sigma = 3600\cdot \frac{180}{\pi}\cdot (a\tan^2{\theta}+b\tan{\theta})
 \end{equation}
Here, $\sigma$ represents the rms blur circle radius [arcsec], and $\theta$ represents the incident angle (the off-axis angle in this paper) [radian]. Assuming that the image blur on the focal plane due to the off-axis imaging characteristics of Wolter-I optics follows a Gaussian distribution with a standard deviation of $\sigma$, HPD is represented as $2\sigma \sqrt{2 \ln2}$. We introduce a constant parameter $c$ to represent the HPD contribution from figure errors and assume that it is independent of the off-axis angle. Assuming independent contributions from the off-axis imaging characteristics of Wolter-I optics and mirror figure errors, the HPD model is expressed as follows.
\begin{equation}
     \mathrm{HPD} = \sqrt{(2\sigma \sqrt{2 \ln2})^2 + c^2}
     \label{eq:HPD_off-axis}
\end{equation}
The data shown in Figure \ref{fig:off-axis_hpd} were fitted with this model. The obtained fitting parameters were $a = 5.6, \, b = 1.8\times10^{-24},\,  c = 12^{\prime\prime}$. 
The model successfully reproduced the overall off-axis angle dependence of the HPD at 12 keV, although deviations of approximately 20\% remained, likely reflecting the azimuthal variations in mirror figure errors discussed above.
The fitted value of $b$ is negligibly small, indicating that the linear term in $\tan\theta$, which would otherwise contribute to the HPD degradation at small off-axis angles, is not required to reproduce the measured dependence. 
This is likely because the image blur near the optical axis is dominated by mirror figure errors, represented by the constant parameter $c$, making the contribution of the linear term difficult to distinguish in the fitting process. 
Consequently, the observed off-axis dependence is primarily described by the quadratic term in $\tan\theta$.
In L. P. VanSpeybroeck and R. C. Chase (1972) \cite{VanSpeybroeck/OPG/1972}, $a$ and $b$ are expressed in terms of the telescope design parameters. Substituting the design parameters of our mirrors yields $a_{\mathrm{design}} = 5.3$ and $b_{\mathrm{design}} = 5.6 \times 10^{-5}$. 
The fitted value of $a$ is comparable to the design prediction, whereas the fitted value of $b$ is substantially smaller. 
The fitted value of $a$ is slightly larger than the design prediction, possibly because stray light enters the detector field of view at large off-axis angles, increasing the measured HPD.
The energy dependence of the on-axis HPD was also evaluated. The HPDs at the best-aligned position for 10, 12, and 15 keV were $14\pm1$, $13\pm1$, and $12\pm1$ arcsec, respectively. These values are consistent within the measurement uncertainties and are also consistent with the fitted parameter $c$ = 12 arcsec and the previous measurement reported by Fujii et al. (2026)\cite{Fujii/PASP/2026}.

\subsection{Effective Area}
\label{subsec:NASA/GSFC}
The results of the effective area measurements at the NASA/GSFC 100 m X-ray beamline are summarized below. The 100 m X-ray beamline at NASA/GSFC consists of an X-ray generator, a monochromator, and optics and detector stages, and is operated at a pressure of $10^{-6}$--$10^{-5}$ Torr\cite{Boissay-Malaquin/SPIE.proc/2022}. The coordinate system was defined as shown in the right panel of Figure \ref{fig:Setup}. The optics were mounted on translation and rotation stages with five degrees of freedom ($y,\,z,\,\theta_x,\,\theta_y,\,$and $\theta_z$), while the detector was mounted on translation stages along the $x$-, $y$-, and $z$-axes. An SDD detector was used for the effective area measurement, and a CCD detector was used for the position and angular alignment. Detailed descriptions of the beamline and the detectors are provided in Boissay-Malaquin et al. (2022)\cite{Boissay-Malaquin/SPIE.proc/2022}.

The optics were aligned in angle and position using the same procedure as described in Section \ref{subsec:SP8}, and the best-aligned position was defined. The effective area was measured using the continuum emission from an unfiltered X-ray source with a Cu target operated at a tube voltage of 14 kV and a tube current of 20 mA. 
With the best-aligned position taken as the reference, effective area measurements were performed at 31 selected angular positions within the range of $-12.5$ to $+12.5$ arcmin in both $\theta_y$ and $\theta_z$, using a grid spacing of 2.5 arcmin.

\begin{figure}[htbp]
\centering
\includegraphics[width=\columnwidth]{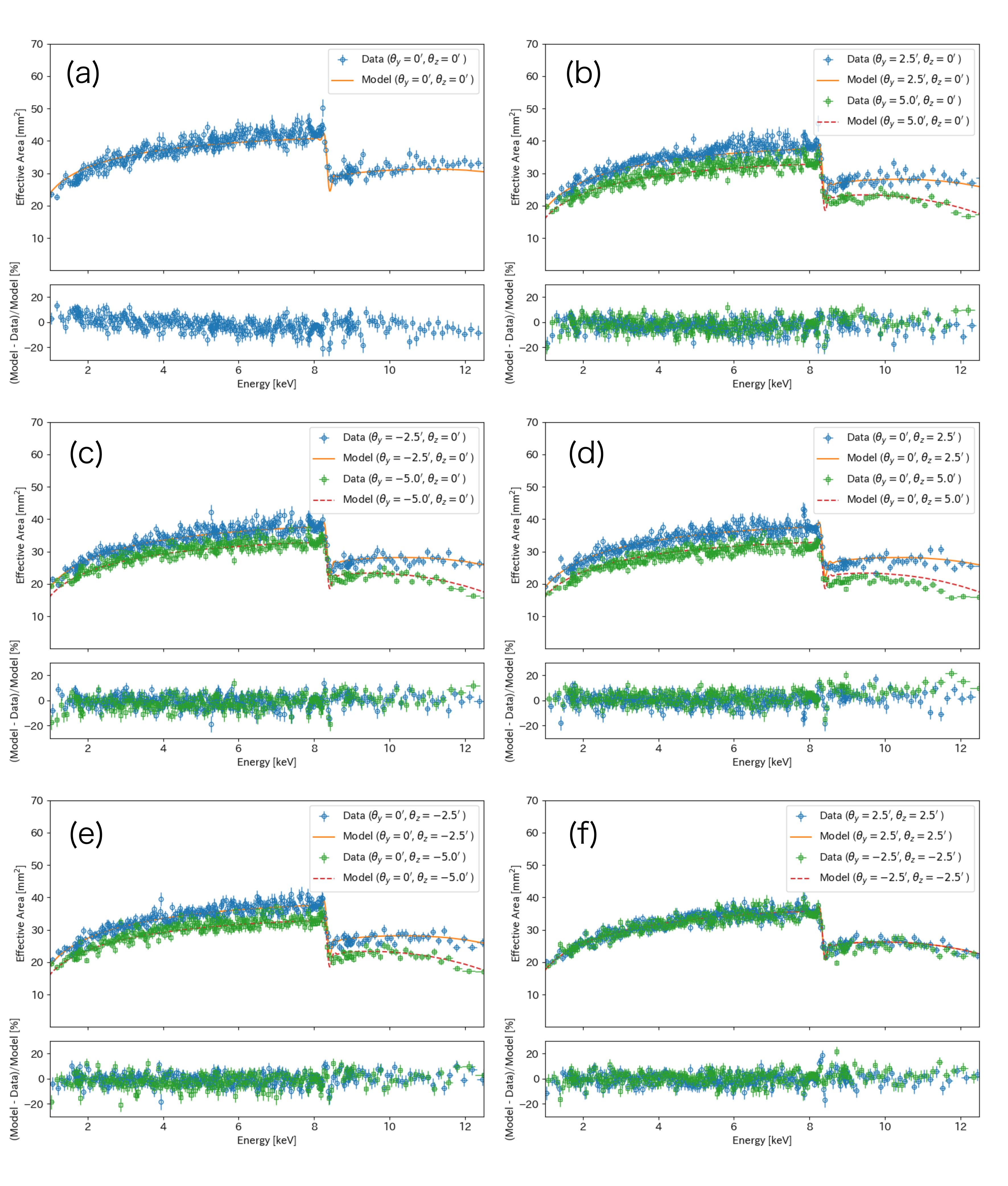}
\caption{Comparison of the measured and modeled effective areas for on- and off-axis rays at the NASA/GSFC 100 m X-ray beamline. The off-axis positions $(\theta_y,\,\theta_z)$ shown in each panel are: (a) $(0',\,0')$; (b) $(2.5',\,0')$ and $(5.0',\,0')$; (c) $(-2.5',\,0')$ and $(-5.0',\,0')$; (d) $(0',\,2.5')$ and $(0',\,5.0')$; (e) $(0',\,-2.5')$ and $(0',\,-5.0')$; and (f) $(2.5',\,2.5')$ and $(-2.5',\,-2.5')$. Measured data are shown as blue circles and green squares, and the corresponding best-fit models are shown as orange solid and red dashed lines, respectively.}
\label{fig:effective_area_data_model}
\end{figure}

The measured effective areas for $\theta \leq 5$ arcmin are shown in Figure \ref{fig:effective_area_data_model}. Here, the off-axis angle is defined as $\theta=\sqrt{\theta_y^2 + \theta_z^2}$. At off-axis angles of $7.5$ arcmin or larger, stray light entered the detector field of view, resulting in an increase in the detected photon counts at lower energies. Consequently, the measured effective area in this range includes a contribution from stray light. The effective area above $12.5$ keV could not be sufficiently evaluated because of the limited number of photons. At $(\theta_y,\,\theta_z) = (0',\,0'),\,(0',\,2.5'),$ and $(0',\,5.0'),$ the effective areas at $6.4$ keV were measured to be $41.3\pm1.9$, $33.7\pm1.7$, and $32.7\pm1.6$ mm$^2$, respectively, indicating an approximately $20\%$ reduction at an off-axis angle of $5.0$ arcmin. In addition, a decrease in the effective area was observed above the Ni K absorption edge at $8.3$ keV. At $10$ keV, the effective areas were measured to be $32.3\pm1.6$, $27.7\pm1.3$, and $21.8\pm1.0$ mm$^2$, respectively, corresponding to a reduction of approximately $30\%$ at an off-axis angle of $5.0$ arcmin. The larger reduction in the effective area for off-axis rays at $10$ keV compared with that at $6.4$ keV is considered to be caused by the incident-angle dependence of the reflectivity. To model the off-axis angle and energy dependence of the effective area, we adopted the approach developed by Ampuku et al. (2024)\cite{Ampuku/SPIE.proc/2024}, in which the effective area is expressed as the product of the mirror effective area, the aperture ratio of the front and rear apertures, the aperture ratio of the pre-collimator, and the transmittance of the optical blocking filter. Since the hard X-ray optics were evaluated in this study, the effects of a pre-collimator and an optical blocking filter are not included. The effective area of the mirror was calculated using a ray-tracing simulation with reflectivities based on the Center for X-Ray Optics database\cite{HENKE/Reftabel/1993}. The aperture ratio was fixed at the design value of approximately $86\%$. However, this model could reproduce neither the overall $\sim20\%$ offset between the calculated and measured effective areas nor the reduction in the effective area below  $6$ keV observed in the measurements. These effects may be attributed to misalignment between the mirror and the rear aperture, surface roughness, reduced density of the reflective surface due to electroforming, oxidation of the reflective surface, and absorption and scattering by contaminant materials. Discrepancies between measured and predicted effective areas, including energy-dependent discrepancies, have been reported for previous X-ray telescopes\cite{Tsusaka/Appl.OPt./1995, Madsen/AAS/2015, Mori/PoS/2018}. We introduced an empirical correction factor to reproduce the overall offset and the low-energy attenuation: $C(E,\theta) = C_0\,\mathrm{exp}(-d\mathrm{(\theta)}\cdot (E_0/E)^{n})$. Here, E is the photon energy [keV], $E_0 = 6$ keV is the reference energy, and $d(\theta)$ is given as a function of the off-axis angle $\theta$ [arcmin], defined as $d(\theta) = d_0 + d_1\theta + d_2\theta^2$. Including this correction factor, the effective area of the hard X-ray optics is modeled as
\begin{equation}
    A_\mathrm{{e, optics}}(E, \theta) = A_\mathrm{{e, mirror}}(E, \theta)\cdot \mathrm{AR}\cdot C(E,\theta)
\end{equation}
Here, $A_\mathrm{{e, optics}}$ and $A_\mathrm{{e, mirror}}$ denote the effective areas of the entire optics and the mirror, respectively, and AR is the aperture ratio. The parameters $C_0,\, d_0,\,d_1,\,d_2,$ and $n$ were determined by fitting all measured data with $\theta \leq 5$ arcmin, where the stray-light contribution is expected to be negligible compared with that of the double reflection component. The best-fit values were $C_0= 0.85,\,d_0 = 0.13,\, d_1= 0.025, \, d_2 = -0.003,\,$and$~n = 0.79$. We successfully reproduced the measured effective areas in the 1.0--12.5 keV energy range at $\theta \leq 5$ arcmin to within $20\%$ (Figure ~\ref{fig:effective_area_data_model}). However, for on-axis rays, the model tends to overestimate the measured effective area at low energies and underestimate it at high energies. The current model makes it difficult to fully reproduce the energy dependence of the effective area, resulting in the observed trend.

To examine the applicability of the empirical model beyond the fitting range, we extrapolated the best-fit model derived from measurements at off-axis angles of up to 5 arcmin to larger off-axis angles without modifying the fitting parameters. 
At an off-axis angle of approximately 7.5 arcmin, the extrapolated model generally reproduced the measured effective areas over the investigated energy range, suggesting that the stray-light contribution remained relatively small under these conditions.
At off-axis angles of 10 arcmin and larger, however, substantial discrepancies appeared, particularly below the Ni K absorption edge (approximately 8.3 keV), where the contribution from stray light becomes increasingly significant. 
In contrast, above the Ni K absorption edge, the measured effective areas were generally reproduced within approximately 20\%, even at larger off-axis angles, although deviations approaching 30\% were observed at the largest off-axis angles, including $(\theta_y,\,\theta_z) = (12.5',\,12.5')$. 
These results suggest that the empirical model can provide a reasonable approximation beyond the fitting range under conditions where the stray-light contribution remains relatively small.

The discrepancies at large off-axis angles and low photon energies are likely attributable to stray-light contributions that are not adequately accounted for by the present model, which primarily describes the double reflection component without explicitly accounting for stray light. 
%
%
Furthermore, the stray-light contribution exhibits a pronounced dependence on azimuthal angle, as indicated by the differences in the measured effective areas at comparable absolute off-axis angles. 
Therefore, extending the effective area model to a wider range of off-axis angles and photon energies may require a separate treatment of stray-light contributions, including their azimuthal dependence.

\section{Summary and Conclusion}
FOXSI-4 and FOXSI-5 are Japan--U.S. joint solar sounding rocket experiments aimed at investigating particle acceleration and plasma heating associated with solar flares.
We provided one soft X-ray optical system and one hard X-ray optical system for FOXSI-4 and FOXSI-5. Since solar flares have extended structures, it is necessary to understand not only the response of the optics to on-axis rays but also the response to off-axis rays. However, although the imaging performance of the hard X-ray optics for on-axis rays had been evaluated in detail, their response to off-axis rays had not been investigated. In this study, we evaluated the imaging performance at SPring-8 BL29XUL to obtain high-quality PSFs, and the effective area at NASA/GSFC over a broad energy range. Focused images were obtained with off-axis angles varied from $-12.5$ arcmin to $+12.5$ arcmin in steps of $2.5$ arcmin, yielding a total of 121 angular positions at $12$ keV. Analysis of these data revealed that, as the off-axis angle increased, the focused image became increasingly elongated and the photon flux decreased. Furthermore, the contribution of the large-angle component increased in the PSF. 
The HPD remained below 20 arcsec for off-axis angles up to 7.5 arcmin, but degraded rapidly at larger off-axis angles, reaching approximately 50 arcsec at 15 arcmin.
To quantitatively describe the off-axis angle dependence of the HPD, we adopted a model that accounts for blurring due to the off-axis imaging characteristics of Wolter-I optics, as well as an on-axis blurring component arising from mirror figure errors, which was assumed to be independent of the off-axis angle.
The model successfully reproduced the overall off-axis angle dependence of the HPD at 12 keV. 

The effective area over the energy range of 1.0--12.5 keV was measured with the off-axis angle varied from $-12.5$ to $+12.5$ arcmin in steps of 2.5 arcmin, yielding a total of 31 angular positions as part of the measurement matrix. In this case, stray light entered the detector field of view at off-axis angles of 7.5 arcmin or greater, so the measured effective area included contributions from stray light. The effective area for on-axis rays was $41.3 \pm 1.9$ mm$^2$ at $6.4$ keV and $32.3 \pm 1.6$ mm$^2$ at $10$ keV. At an off-axis angle of 5 arcmin, the effective area decreased by 20\% at $6.4$ keV and 30\% at $10$ keV, indicating that the degree of reduction also depends on energy. The measured effective areas were generally about 20\% lower than those calculated by ray-tracing simulations, with a tendency for the discrepancy to increase toward lower energies. To account for this discrepancy, we introduced an empirical correction factor and fitted it to the data obtained within an off-axis angle of 5 arcmin. The results confirmed that the model reproduced the off-axis angle dependence of the effective area to within 20\% for off-axis angles up to 5 arcmin.
Extrapolation to larger off-axis angles showed reasonable agreement at approximately 7.5 arcmin and generally within 20\% above the Ni K absorption edge, even at larger off-axis angles. 
However, substantial discrepancies were observed at low energies for off-axis angles of 10 arcmin and larger, where stray-light contributions become significant. 
The pronounced azimuthal dependence of the stray-light contribution suggests that further refinement of the effective area model may require explicitly accounting for this component and its azimuthal dependence.

The hard X-ray optics characterized in this study were successfully flown on FOXSI-4 in 2024 and FOXSI-5 in 2026. 
The off-axis imaging performance and effective area measurements presented here provide essential calibration data for constructing mission-specific response functions, taking into account the alignment of each mission.

\subsection*{Disclosures}
The authors declare that there are no financial interests, commercial affiliations, or other potential conflicts of interest that could have influenced the objectivity of this research or the writing of this paper.

\subsection* {Code, Data, and Materials Availability} 
The datasets generated and/or analyzed during the current study are available from the corresponding author on reasonable request.

\subsection* {Acknowledgments}
The FOXSI-4 and FOXSI-5 sounding rocket missions were funded by the NASA Low Cost Access to Space (LCAS) Program under Grant Numbers 80NSSC22K0577 and 80NSSC25M7045, respectively. 
We acknowledge the National Science Foundation (NSF) for its support of space physics at the University of Minnesota through Grants AGS-1429512 and AGS-1752268. 
The FOXSI-4 mission was also supported by the ISAS/JAXA Small Science Program and JSPS KAKENHI under Grant Number JP21H04486. 
The FOXSI team is grateful to the NASA and NSROC teams at Wallops and White Sands Missile Range (WSMR), and to the staff of Poker Flat Research Range and the University of Alaska Fairbanks (UAF) Geophysical Institute, for supporting both missions. 
We also acknowledge the contributions of all members of the FOXSI experiment team across the many participating institutions, areas of expertise, and career levels.
This work was supported by JSPS KAKENHI Grant Numbers JP22K18274 and JP20K20920 (I.M.), JP23H00156 (H.M.), and JP18H03724, JP22H00134, and JP21KK0052 (N.N.), as well as by the JST FOREST Program (Grant Number JPMJFR2369, Japan) (I.M.).
Additional support was provided by JST SPRING (Grant Number JPMJSP2125). 
This work was carried out under the Joint Research Program of the Institute for Space--Earth Environmental Research (ISEE), Nagoya University. The synchrotron radiation experiments were performed at BL29XUL of SPring-8 with the approval of the Japan Synchrotron Radiation Research Institute (JASRI) (Proposal No. 2024A1457). The use of BL29XUL at SPring-8 was supported by RIKEN. 
Y.Y. and K.S. acknowledge support from the JSPS Research Fellowships for Young Scientists (DC1 and PD, respectively) under Grant Numbers JP26KJ1366 and JP26KJ0023, respectively.
Y.Y. also acknowledges financial support from the Hayakawa Satio Fund awarded by the Astronomical Society of Japan. 
M.Y. acknowledges financial support from the Iwadare Scholarship Foundation. K.A. acknowledges the generous support of the Hattori International Scholarship Foundation (HISF). 
K.S. and K.A. also acknowledge financial support from the ``THERS Make New Standards Program for the Next Generation Researchers.'' 
OpenAI's ChatGPT and Anthropic's Claude were used to assist with English editing, translation, and grammar support.


\bibliography{report}   
\bibliographystyle{spiejour}   


\vspace{2ex}\noindent\textbf{Yusuke Yoshida} is a PhD student at Nagoya University's Uxg Lab. He earned bachelor's and master's degrees from Nagoya University in 2024 and 2026. His research interests include X-ray telescope development. He is a member of SPIE and the Astronomical Society of Japan. He received a poster award at SPIE Astronomical Telescopes + Instrumentation 2024. He works on FOXSI-4/5 X-ray optics, including ground calibration and modeling. He is developing X-ray instrumentation for CubeSats and microsatellites.

\vspace{2ex}\noindent\textbf{Ikuyuki Mitsuishi} received his PhD in physics from the University of Tokyo in 2012. He is currently a lecturer at the Graduate School of Science, Nagoya University. He led the development of thin-film filters for the FOXSI-3 sounding rocket experiment and thermal shields for the Imaging X-ray Polarimetry Explorer (IXPE), as well as the Japanese-provided soft and hard X-ray optics for the FOXSI-4 and FOXSI-5 sounding rocket missions. He is currently the principal investigator of COMIX (Compact Modular Imager for X-rays), a project developing ultrashort-focal-length X-ray optics and ultrathin graphene-based filters for small space platforms, including CubeSats and microsatellites. His observational research interests span high-energy astrophysical phenomena across a wide range of spatial scales, from stars and the Milky Way to starburst galaxies, galaxy clusters, and large-scale structures in the Universe.

\vspace{1ex}
\noindent Biographies of the other authors are not available.

\listoffigures
\listoftables

\end{spacing}
\end{document}